\documentstyle[preprint,aps,eqsecnum]{revtex}
\def\beq{\begin{equation}}
\def\eeq{\end{equation}}
\def\ra{\rightarrow}
\begin{document}

{\tighten
\preprint{\vbox{
\hbox{FERMILAB--PUB--97/384--T}}}
\title{\bf Clean CKM Information from $B_d (t) \rightarrow D^{(*)\mp} \pi^\pm$}
\author{Isard Dunietz}
\address{\it Fermi National Accelerator Laboratory, P.O. Box 500,  
Batavia, IL
60510}

\bigskip
\date{\today}
\maketitle
\begin{abstract}
It has been known for many years that the $B_d (t) \rightarrow D^{(*)\mp
}\{\pi^\pm ,\rho^\pm ,a_1^\pm\}$ modes may involve observable CP violating effects.
This note describes how to determine cleanly the Cabibbo-Kobayashi-Maskawa (CKM) phase $\phi=-2\beta
-\gamma=-\pi +\alpha -\beta$, even in the presence of possible final state interactions. A discrete ambiguity remains.

\end{abstract} }
\newpage
\section{Introduction}
The next decade will witness an unprecedented number of consistency checks on
whether the CKM (Cabibbo-Kobayashi-Maskawa) hypothesis~\cite{ckm} correctly describes CP violation. While the gold-plated
$B_d\ra J/\psi K_S$ asymmetry~\cite{bigisanda} cleanly determines $\sin 2\beta$, the other angles
of the CKM unitarity triangle are harder to obtain \cite{burasfleischer}.

Here we
report on a clean method that extracts the CKM phase combination $2\beta +\gamma$
or $\beta -\alpha$.  While in principle all hadronic uncertainties can be disentangled~\cite{adk,silva}, in practice this is unfeasible for first generation experiments.  Those experiments could determine the CKM phase, however, by incorporating related modes, as shown below. Since $\beta$ will be known, the CKM angle $\alpha$ or
$\gamma$ could be obtained cleanly [up to a discrete ambiguity].
This is important, because accurate knowledge of the CKM parameters will constrain
or rule out the CKM explanation for CP violation.

A time dependent study of $B_d
(t) \ra \pi^+ \pi^-$ is not capable of extracting the CKM parameter $\sin 2\alpha$
cleanly, because of non-negligible penguin amplitudes. That can be inferred from the recent CLEO result for
$\pi^+\pi^- /K^+ \pi^-$~\cite{cleokpi}, which indicates that penguin amplitudes are sizable in $B\ra
\pi\pi$ transitions. Thus, the clean determination of $\alpha$ from $B\ra \pi\pi$
modes requires the study of $B_d \ra \pi^0\pi^0$ \cite{isospin} which is almost
impossible at hadron accelerators [see, however, Ref.~\cite{snowmass}], but may be
possible at $\Upsilon (4S)$ factories if the branching-ratio is not too small.
Quinn and Snyder proposed to determine $\alpha$ from Dalitz plot analyses of $B_d
\ra \rho\pi$ \cite{quinnsnyder}.
That method works if the non-resonant and other $B\ra 3\pi$ amplitudes are well
understood, which may require large statistics \cite{lediberder}. Those modes
involve $\pi^0$'s and thus can be more naturally studied at present at $e^+e^-$
colliders. Experiments at hadron accelerators would greatly enhance their
$b$-physics reach by developing methods for efficient photon, $\pi^0 , \eta,\eta^\prime$ reconstruction.

Denote the $B_d /\overline B_d$ modes $D^{(*)-}
\{\pi^+ ,\rho^+, a^+_1 ,...\}$ by $f$, and $\overline f \equiv$ CP $f$. Sachs stressed the importance of such
non-leptonic, non-CP eigenstates in mixing-induced CP violation studies
\cite{sachs}. Until then, mixing-induced CP studies focused on either
same-sign dilepton asymmetries \cite{dilepton} or on CP eigenmodes
\cite{bigisanda,wolfenstein}.
However, CP violation can also be seen [either time-dependent or
time-integrated] with non-CP eigenstates \cite{sachs,dunietzr,ddw}:
\beq
\label{smallcp}
\Gamma (B_d (t) \ra f) \neq \Gamma (\overline B_d (t) \ra \overline f) \;,
\eeq
\beq
\label{largecp}
\Gamma (B_d (t) \ra \overline f) \neq \Gamma (\overline B_d (t) \ra f) \;.
\eeq

For instance, the $B_d (t) \ra \overline f$ process involves the direct amplitude $B_d
\ra \overline f$ governed by the tiny CKM combination $V_{ub}^* V_{cd}$ and the
mixing-induced amplitude $B_d(t)\ra \overline B_d \ra \overline f$, where the latter
$\overline B_d\ra \overline f$ transition is governed by the CKM favored $V_{cb}
V^*_{ud}$ combination.\footnote{Those CKM combinations are unique and the same for either
the color-allowed spectator graph or the internal-$W$ graph, predicted to
be much smaller.} The disparate strengths of the two interfering amplitudes cause 
the CP asymmetry to be at the few percent level.
The CP asymmetry is larger for process (\ref{largecp}) than for process
(\ref{smallcp}), because the two interfering amplitudes are made significantly less
disparate in size by the judicious positioning of the small $B_d -\overline B_d$
mixing-amplitude.

Since the distinction between an initially unmixed $B_d$ and $\overline B_d$ (flavor-tagging) entails normally some impurity, one will have to correct
for a (serious) asymmetric background $[\overline B_d (t)\ra \overline f]$.  The correction is, however, well
understood because it depends on the same observables (see
Eq.~(\ref{observables}) below).

The interference term is \cite{adk}
\beq
\lambda \equiv \frac{q}{p} \;\frac{\langle f|\overline B_d\rangle}{\langle f|B_d
\rangle} =\rho \;e^{i(\phi+\Delta )} \;,
\eeq
where $\rho$ denotes the magnitude of the amplitude ratio. The weak phase (CKM
phase) is
\beq
\phi =-2\beta -\gamma =-\pi +\alpha -\beta \;,
\eeq
and $\Delta$ denotes a possible strong phase difference.
Ref.~\cite{adk} demonstrated how a time-dependent study of the four rates,
$$\Gamma (B_d (t) \ra f), \Gamma (B_d (t) \ra \overline f), \Gamma (\overline B_d (t)
\ra \overline f), \Gamma (\overline B_d (t) \ra f),$$
extracts the three observables\footnote{For a non-vanishing $B_d - \overline B_d$ width difference $\Delta \Gamma$ (expected to be at the $1\%$ level~\cite{bbd}), the relevant observables can be extracted from fits to more involved time-dependences~\cite{bsbsbar}.  Of course, the accurate extraction of the CKM angle $\beta$ will also involve more elaborate fits.}
\beq
\label{observables}
\rho , \;\sin (\phi +\Delta),\;\sin (\phi -\Delta )\;.
\eeq
The weak phase $\phi$ can be determined up to a discrete ambiguity from
fundamental trigonometry. 
By the time such demanding studies can be performed, the angle $\beta$ will be
well known from the $B_d \ra J/\psi K_S$ asymmetry. Thus the angle $\alpha$ (or
$\gamma$) can be \underline{cleanly} extracted, because
penguin amplitudes cannot contribute. A discrete ambiguity remains.\footnote{It maybe partially resolved because $\phi$ is the same for the various modes, whereas $\Delta$ could be mode-dependent.}

While our observations are true in principle, it is exceedingly difficult, in
practice, to fit for such small $\rho$ parameters in time-dependent studies. We
therefore suggest to determine $\rho$ elsewhere.  The observable $\rho^2$ is essentially the ratio of rates,
$$\rho^2 = \frac{\Gamma (\overline B_d \ra D^{(*)-} \{\pi^+ , \rho^+, a_1^+,
...\})}{\Gamma (B_d \ra D^{(*)-} \{\pi^+ , \rho^+, a_1^+,
...\})} \;.$$
The difficulty in obtaining $\rho^2$ lies in determining the tiny numerator.  That numerator can be obtained by studying the (a) $\overline B_s \to D^{(*)-} K^{(*)+}$ processes, (b) $\overline B_d \to D_s^{(*)-} \{\pi^+ , \rho^+, a_1^+,
...\}$ processes [where the strangeness content of the final state automatically tags the $\overline B$ flavor at time of decay], (c)
$$\rho^2 \approx \frac{2\;\Gamma (B^- \ra D^{(*)-} \{\pi^0 , \rho^0, a_1^0,
...\})}{\Gamma (B_d \ra D^{(*)-} \{\pi^+ , \rho^+, a_1^+,
...\})} \;.$$
Small corrections to the approximations can be incorporated once they have
been investigated experimentally and theoretically \cite{rosnerdpi}.

Furthermore, it is probable that the strong phase difference is small $\Delta
\approx 0$ mod $\pi$, \footnote{Both interfering amplitudes are governed dominantly
by color-allowed spectator graphs.  Under the factorization approximation no final
state phase difference occurs. Such phase differences could be generated by
non-factorizable contributions, rescatterings and subleading diagrams as catalogued
in Ref.~\cite{rosnerdpi}, and are all expected to be small.} and a first generation
experiment may wish to fit the four time-dependent rates for the single parameter
$\phi$. To determine $\sin \phi$ to an accuracy of $\pm 0.1$, one requires about $10^8$ (fully) flavor-tagged $B_d + \overline B_d$ mesons.  This estimate assumes ${\cal O} (1)$ detection efficiencies and a nominal $B(B_d \to f) = 0.003$. The relevant branching-ratios $B(B_d \to f)$ are indeed large ${\cal O} (0.003 - 0.01)$~\cite{pdg96}.  The main difficulty lies in accurately observing the small asymmetry governed by the interference term $\lambda$, whose magnitude is approximately
$$\rho \approx |V_{ub}/V_{cb}| \times |V_{cd}/V_{ud}| \approx 0.02.$$
We anticipate that the observable $\rho^2$ will be known to sufficient accuracy, because several $\rho^2$ extractions do not require flavor-tagging and could infer $\rho^2$ from less CKM suppressed transitions (as outlined above).
 
The unique kinematics of the $D^{*\pm} \pi^\mp$ mode may permit a semi-inclusive
reconstruction by using the soft $\pi^\pm$ in the $D^{*\pm}\ra\pi^\pm
\stackrel{(-)}{D^0}$ decay \cite{nelson}. Those modes are also a natural for
detectors operating at hadron accelerators, because the analogous $B_s\ra D^-_s \pi^+,
D^-_s \pi^+\pi^- \pi^+$ processes, important for $\Delta m_s$ measurements, were
shown to be accessible with large rates.
Further note that the $B_d\ra D^{*\mp} \{\rho^\pm, a^\pm_1\}$
modes allow also the clean determination of the CKM phase $\phi$, by employing
angular correlations. Those angular correlations permit the study of more involved CP violating observables.

In conclusion, this note demonstrates that the $B_d\ra D^{(*)\mp} \{\pi^\pm
,\rho^\pm ,a_1^\pm ,...\}$ transitions allow the clean extraction of the CKM
angle $\phi =-2\beta -\gamma =-\pi +\alpha -\beta$. That may play a role in
constraining or ruling out the CKM hypothesis for CP violation.

\section{Acknowledgements}

We are grateful to R. Fleischer and C. Quigg for informing us that the BABAR collaboration is studying the feasibility of such modes by semi-inclusive reconstruction. We thank J. Lewis, J.L. Rosner, M. Shapiro and S. Stone for discussions. This work was supported in part by the Department of Energy, Contract No.
DE-AC02-76CH03000.

\end{document}